\documentclass[conference]{IEEEtran}
\IEEEoverridecommandlockouts
\usepackage{cite}
\usepackage{amsmath,amssymb,amsfonts}
\usepackage{algorithmic}
\usepackage{graphicx}
\usepackage{textcomp}
\usepackage{multirow}
\usepackage{xcolor}
\usepackage{booktabs}
\usepackage{threeparttable}

\def\BibTeX{{\rm B\kern-.05em{\sc i\kern-.025em b}\kern-.08em
    T\kern-.1667em\lower.7ex\hbox{E}\kern-.125emX}}

\usepackage{tikz}
\usepackage[hyperfootnotes=false]{hyperref} 

\newcommand\copyrighttext{%
  \footnotesize \textcopyright 2021 IEEE. Personal use of this material is permitted.
  Permission from IEEE must be obtained for all other uses, in any current or future
  media, including reprinting/republishing this material for advertising or promotional
  purposes, creating new collective works, for resale or redistribution to servers or
  lists, or reuse of any copyrighted component of this work in other works.
  DOI: \href{https://doi.org/10.1109/IPCCC51483.2021.9679361}{https://doi.org/10.1109/IPCCC51483.2021.9679361}}
\newcommand\copyrightnotice{%
\begin{tikzpicture}[remember picture,overlay]
\node[anchor=south,yshift=10pt] at (current page.south) {\fbox{\parbox{\dimexpr\textwidth-\fboxsep-\fboxrule\relax}{\copyrighttext}}};
\end{tikzpicture}%
}

\newcommand*\concat{\mathbin\big\Vert}
\DeclareMathAlphabet{\altmathcal}{OMS}{cmsy}{m}{n}
    
\begin{document}

\title{Enel: Context-Aware Dynamic Scaling of Distributed Dataflow Jobs using Graph Propagation}

\author{
\IEEEauthorblockN{Dominik Scheinert, Houkun Zhu, Lauritz Thamsen, Morgan K. Geldenhuys,\\Jonathan Will, Alexander Acker, and Odej Kao}
\IEEEauthorblockA{Technische Universit{\"a}t Berlin, Germany, \{firstname.lastname\}@tu-berlin.de}
}

\maketitle
\copyrightnotice

\begin{abstract}
Distributed dataflow systems like Spark and Flink enable the use of clusters for scalable data analytics. 
While runtime prediction models can be used to initially select appropriate cluster resources given target runtimes, the actual runtime performance of dataflow jobs depends on several factors and varies over time. 
Yet, in many situations, dynamic scaling can be used to meet formulated runtime targets despite significant performance variance.

This paper presents \emph{Enel}, a novel dynamic scaling approach that uses message propagation on an attributed graph to model dataflow jobs and, thus, allows for deriving effective rescaling decisions. For this, Enel incorporates descriptive properties that capture the respective execution context, considers statistics from individual dataflow tasks, and propagates predictions through the job graph to eventually find an optimized new scale-out. Our evaluation of Enel with four iterative Spark jobs shows that our approach is able to identify effective rescaling actions, reacting for instance to node failures, and can be reused across different execution contexts.
\end{abstract}


\section{Introduction}
\label{sec:introduction}

Across all domains the amount of data being generated is continually increasing and consequently the need to process and analyse large volumes of data has become an increasingly essential part of any data processing environment.
It is here where scalable data-parallel processing jobs are frequently employed to extract valuable information from this data.
Nowadays these processing jobs can be developed with the help of distributed dataflow systems like Spark~\cite{Zaharia2010} and Flink~\cite{Carbone2015}, which take care of the parallelism, distribution, and fault tolerance.
As the management of computational resources is often not directly handled by these distributed dataflow systems, they commonly make use of resource management systems such as YARN~\cite{Vavilapalli2013} or Kubernetes~\cite{BorgKubernetes2015} to provision resources for the individual processing jobs.

Yet, as the configuration of clusters is traditionally a manual task, finding near optimal resource allocations is difficult as even frequent users and experts do not always fully understand system and workload dynamics~\cite{Rajan2016,Lama2012}.
As a consequence, providing methods which automate this process has become a popular area of research~\cite{Verma2011,Lama2012,Rajan2016,Hsu2018,Casimiro2019,Fekry2020}. 
The need for more sophisticated solutions has been further reinforced by the rise of cloud computing which has led to an increase in the diversity of resource and infrastructures, as well as, users data processing requirements~\cite{Bux2013,Deelman2019}.
In light of this situation, it is necessary to support users in finding a suitable resource configuration that is in line with their respective requirements. 

Various solutions have been proposed regarding the optimization of resource configurations.
Some methods are designed for specific processing frameworks~\cite{Verma2011a,Ferguson2012,AlSayeh2019}, others conduct an iterative profiling strategy~\cite{Alipourfard2017,Hsu2018,Hsu2018a,Hsu2018b}, and a third line of work builds runtime models for evaluating possible configurations\cite{Verma2011,Venkataraman2016,Thamsen2016,Thamsen2017,Shah2019,Will2021c3o,scheinert2021bellamy}. 
However, even with optimized initial resource allocation, the performance of processing jobs can vary greatly due to factors such as data locality, inference between jobs, and failures which impact the performance over time. 
Consequently, processing jobs should be continuously monitored to ensure any QoS (Quality of Service) requirements involving job runtimes and resource utilization should be adhered to.
At the same time, it is desirable to develop reusable solutions in order to reduce the initial expense of finding near-optimal configurations.

In this paper we present \emph{Enel}, a black-box approach to continuously optimizing processing jobs where the underlying graph structure of data-parallel tasks are explicitly taken into consideration.
In contrast to related state of the art methods~\cite{Koch2017,Thamsen2017}, our approach models a distributed dataflow job and its components as an attributed, directed graph and aims to learn the relationship between nodes and their respective predecessor nodes.
This is fostered by attaching descriptive properties to the individual nodes that at best describe the respective context of the job execution, e.g. dataset size, algorithm name, or machine type. 
The learned relationships and attached context information are then used to predict the individual runtimes and runtime statistics.
During online inference, data emitted from the current job execution is gathered and used to continuously update runtime predictions. 
Based on a specified runtime target and the remaining predicted runtime, the current resource allocation is eventually optimized to bear in mind QoS constraints.

\emph{Contributions.} The contributions of this paper are:

\begin{itemize}
    \item A novel modeling approach for dynamic scaling that incorporates scale-out information, descriptive properties of the job execution context, and application metrics, while also exploiting the underlying graph structures of data-parallel tasks in distributed dataflow jobs. We show that hereby, we can achieve accurate and robust prediction results and thus a reliable dynamic scaling mechanism.
    \item An evaluation of our approach to dynamic scaling in various different experiments. We investigate our approach in the presence of failures, and show that our graph model is reusable in different contexts.
    \item A repository\footnote{https://github.com/dos-group/enel-experiments} with all relevant source code and data, accompanied by comprehensive information regarding setups, configurations, and application of our method.
\end{itemize}

\emph{Outline}. The remainder of the paper is structured as follows.
\autoref{sec:related_work} discusses the related work.
\autoref{sec:approach} describes and discusses our modeling approach. 
\autoref{sec:architecture} outlines the architecture decisions behind Enel.
\autoref{sec:evaluation} presents the results of our comprehensive evaluation. 
\autoref{sec:conclusion} concludes the paper and gives an outlook towards future work.
\section{Related Work}
\label{sec:related_work}

Existing solutions for dynamic resource allocation problems of distributed data processing jobs typically involve choosing an initial configuration offline and subsequently adjusting the resources for the running job online, i.e. during the execution.

This section discusses related work that addresses both of these sub-problems and puts them in relation to Enel.

\subsection{Offline Runtime Prediction}

Several model-based offline approaches for selecting cluster configurations exist.
They often use runtime data to predict the scale-out and runtime behavior of jobs.
The runtime data can be gained either from dedicated profiling with a sample of the dataset or previous full executions~\cite{Venkataraman2016,bilal2020finding ,Thamsen2016, Will2020towards,Will2021c3o,scheinert2021bellamy}.

\emph{Ernest}~\cite{Venkataraman2016} uses a parametric model for the scale-out behavior of jobs, which is trained on the results of sample runs on reduced input data.
This succeeds for programs which exhibit intuitive scale-out behaviors.
Ernest uses optimal experiment design to select configurations to test in the sample runs.

Another example is \emph{Vanir}\cite{bilal2020finding}, which firstly finds a good-enough initial configuration for profiling runs by using a heuristics method, and then utilizes Mondrian forest based performance model and transfer learning to progressively improve the configuration in the production runs. In addition, Vanir employs transfer learning in the profiling phase to eliminate the cost of profiling runs for similar jobs.

\emph{Bell}~\cite{Thamsen2016} is an approach which can continue to learn the job's scale-out behavior from historical full executions besides initial runs. It automatically chooses via cross-validation between a parametric model based on that of Ernest and a non-parametric model, which leads to an overall increase in prediction accuracy.
Enel uses Bell's runtime model as part of its process of finding a good initial resource allocation.

Gaining training data for the profiling leads to overhead in both time and to some extent also cost.
Meanwhile, metrics from previous executions of a job are not always available.
As a possible remedy for this issue, we previously proposed a system~\cite{Will2020towards} that facilitates the global sharing of context aware runtime models, allowing for runtime prediction based on historical executions of a job by different users~\cite{Will2021c3o,scheinert2021bellamy}. 
Enel assumes a recurring job and thus, the initial profiling cost can be amortized over time.
However, even the most well-trained runtime models cannot anticipate failures in the cluster, which could endanger reaching the runtime target. 
Likewise, data locality will also have an impact on the runtimes of frequently performed jobs as tasks are not guaranteed to be scheduled on the same nodes where the data files reside. 
Enel therefore additionally monitors the execution and dynamically scales the cluster as required to reach runtime targets. 

\subsection{Progress Estimation and Dynamic Scaling}

There are several works that attempt to estimate the progress of iterative distributed data processing jobs~\cite{sahni2015cost,thonglek2021auto, Thamsen2017}. These estimates can then be used to continuously re-evaluate the current resource configuration and make adjustments towards a given runtime target for the job~\cite{Thamsen2017}.
Besides meeting deadlines, progress estimation can also be used to support scheduling decisions in shared clusters~\cite{Renner2016,Xu2016,Shi2021}.

Sahni and Vidyarthi~\cite{sahni2015cost} propose a dynamic and cost-effective algorithm to schedule deadline-constrained jobs in public clouds. It incorporates information on performance variability among virtual machines and instance acquisition delay. To achieve the cost-effective goal, it scales in a just-in-time schedule manner.

Thonglek and et al.~\cite{thonglek2021auto} utilizes model-based deep reinforcement learning (RL) to achieve an auto-scaling system for real time processing applications. To address the problem of dynamic input data size and unpredictable availability of resources, it applies deep RL to continuously optimize scaling by observing and learning from the previous settings.

\emph{Ellis}~\cite{Thamsen2017} uses progress estimates as a basis for dynamic re-allocations of resources to meet a runtime target.
Progress is estimated by continuously monitoring scale-out behavior of the job's individual stages and comparing it to that of previous executions. It predicts the remaining runtime of jobs and then scales out to meet the runtime target if needed.

Most previous works either utilize no information on resource availability and cluster loads, or do not consider the directed acyclic graph (DAG) of jobs, which makes it harder to satisfy user constraints in complex environments, e.g, running in a multi-tenant cluster. Enel, however, uses message propagation on an attributed graph to model the progress of distributed dataflow jobs and only uses a single global model, thus being more robust to complex situations like resource competitions and small changes of the execution context.

\section{Approach}
\label{sec:approach}

This section presents our approach Enel and its usage to select resources according to formulated runtime targets. As it is designed for reusability, context-awareness, and operates on graphs, we will discuss the individual aspects in the following.

\subsection{Preliminaries}
\label{sec:approach_preliminaries}
\begin{figure}[h]
    \centering
    \includegraphics[width=\columnwidth]{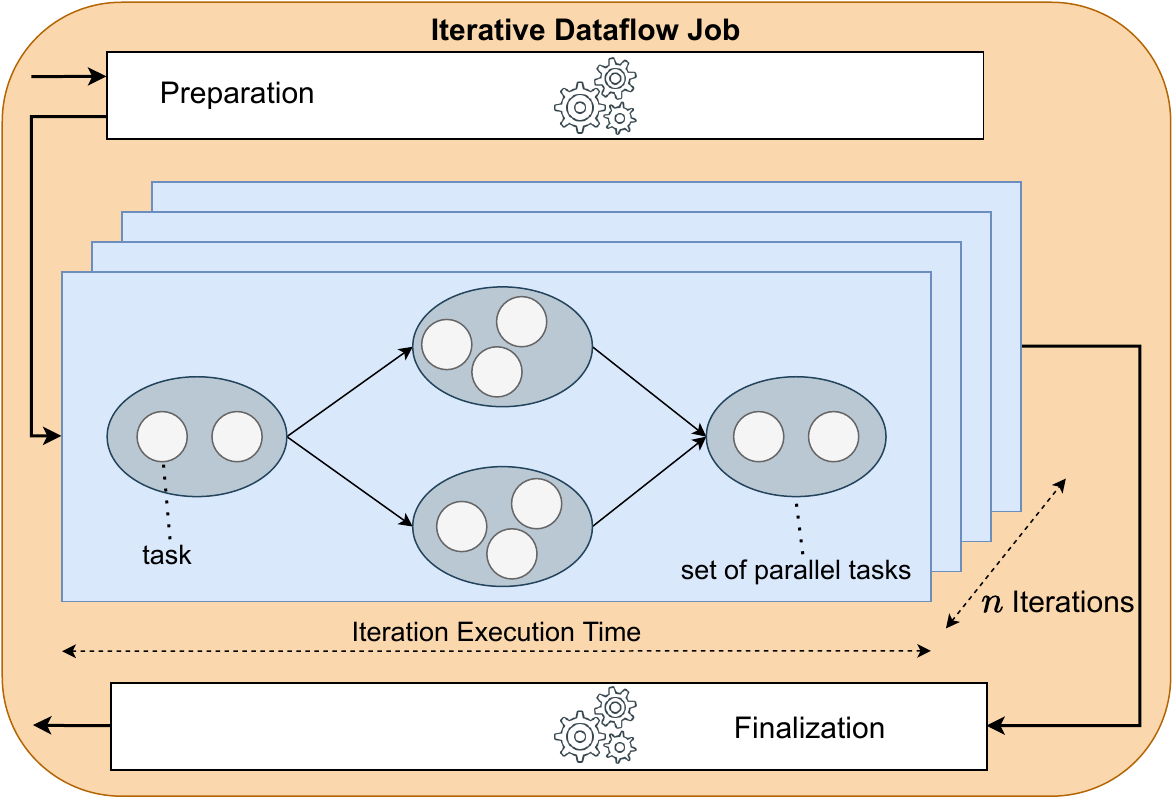}
    \caption{An iterative distributed dataflow job. Typically, an initial preparation phase is followed by a number of actual iterations of an algorithm, with a finalization phase eventually concluding the computation.}
    \label{fig:approach_iterative_dataflows}
\end{figure}

Distributed dataflows are commonly modeled as graphs to represent the temporal relationships between consecutive or parallel actions.  
A directed, weighted, and attributed graph $G^{(k)}=(V^{(k)},E^{(k)})$ consists of a set of vertices $V^{(k)}=\{v_1, \ldots, v_n\}$ and a set of edges $E^{(k)}\subseteq \{(v_i,v_j)| v_i,v_j \in V^{(k)}\}$. 
One can also write $V(G^{(k)})$ and $E(G^{(k)})$.
An edge $e_{ij} \Leftrightarrow (v_i,v_j)\in E^{(k)}$ describes a directed connection between vertex $v_i$ and $v_j$, and $|e_{ij}|$ is the corresponding edge weight. Thus, the node $v_j$ is then called a neighbor of node $v_i$, formally written as $j\in \altmathcal{N}(i)$. Throughout this paper, we use the notion of $G^{(k)}$ to refer to the $k$-th component of a distributed dataflow job, e.g. a concrete iteration in the case of iterative dataflows. 
Consequently, we define a dataflow job as a sequence $D=(G^{(1)}, G^{(2)}, \ldots, G^{(n-1)}, G^{(n)})$ of $n$ graphs, each corresponding to a component of the job.
This is depicted in~\autoref{fig:approach_iterative_dataflows} and is of importance to our approach. 

The execution of distributed dataflows is characterized by descriptive properties. These are different based on the concrete dataflow system and execution environment. 
Examples are job parameters, hardware specifications of the underlying infrastructure, the size of the dataset to be processed, or versions of utilized software.
Properties are furthermore available on multiple levels, i.e. a node $v_i\in V(G^{(k)})$ is characterized by different properties than the corresponding component $k$. We estimate the \emph{execution context} of a job using the entirety of its descriptive context properties.

\subsection{General Idea}
\label{sec:approach_general_idea}
While an initial resource allocation for a target runtime can be computed using historical data or profiling, there are a variety of factors influencing the actual runtime, e.g. data locality or failures. 
In order to meet given runtime targets despite these factors, it is advisable to continuously monitor the progress of a dataflow job and to make changes in the resource allocation, when jobs do not progress as expected.

\begin{figure}[h]
    \centering
    \includegraphics[width=\columnwidth]{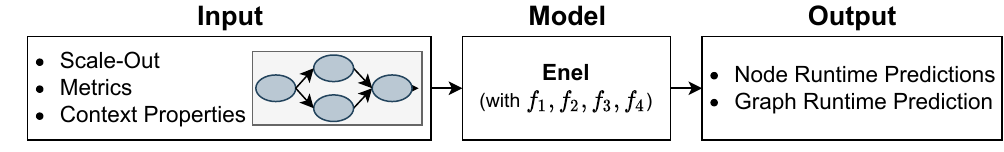}
    \caption{High-level overview of approach for a single graph.}
    \label{fig:approach_overview}
\end{figure}

With our approach \emph{Enel}, we aim at learning to predict the scale-out behavior of jobs by explicitly incorporating the job graph for modeling distributed dataflows.
As each graph node is described by its descriptive context properties and also runtime statistics, we are enabled to make predictions based on locally available information and the respective predecessor nodes. 
Estimated runtimes and metrics are eventually propagated through the graph to successor nodes, which in turn use them to compute own predictions.
Runtime predictions are accumulated in the process of propagation, providing a runtime prediction for the respective graph, as illustrated in~\autoref{fig:approach_overview}.

For training our graph-based model, it is required to present it with a sufficient amount of graphs. 
Therefore, our approach is especially suitable for iterative dataflows as the number of available graphs to learn from is numerous, since each iteration results in a new component modeled as graph. Later, during monitoring of a job execution, our trained graph model can be used to predict the remaining runtime of the job, and potentially adapt the scale-out to meet the given runtime target.

\subsection{Context Encoding}
\label{sec:approach_context_encoding}
In order to make use of descriptive properties of a job execution context, we require an efficient, yet robust way of representing these properties. 
In a first step, we transform each property $p$ to a vector of fixed-size length $\Vec{p}\in\mathbb{R}^N$, i.e.

\begin{equation}
\Vec{p} = [\lambda, \Vec{q}_1, \Vec{q}_2, \ldots, \Vec{q}_{L-1}, \Vec{q}_L]^\top,
\end{equation}

where $\Vec{q}\in \mathbb{R}^L$ with $L=N-1$ is a vector and $\Vec{q}_1,\ldots,\Vec{q}_L$ its elements, obtained from an encoding method as 

\begin{equation}
\Vec{q} = \begin{cases}
\text{hasher}(p) & p \not\in \mathbb{N}_0\\
\text{binarizer}(p) & p \in \mathbb{N}_0

\end{cases}
\end{equation}

and $\lambda \in \{0,1\}$ is a binary prefix indicating the utilized method, i.e. hasher or binarizer.

The \textit{hasher} method operates on textual properties (e.g. job parameters) and combines character cleansing and n-gram extraction to derive terms. 
The occurrence of each resulting term $t_s$ is then counted and inserted at a specific position in the output vector, such that $\Vec{q}_{j} = |t_s|$, where the index $j$ is calculated by a hash function that realizes the term to index mapping.
Lastly, the resulting vector $\Vec{q}$ is projected on the euclidean unit sphere, such that $\sum_{j=1}^L(\Vec{q}_j)^2 = 1$ is ensured. 

The \textit{binarizer} method takes a natural number and converts the respective value into its binary representation. 
As a consequence, each property $p \in \mathbb{N}_0$ (e.g. number of CPU cores) can be encoded as long as $p \leq 2^{L}$ holds true.
This allows for uniquely encoding any number of reasonable size. 

As many of the hereby computed vectors can be expected to be sparse, we further employ an auto-encoder to obtain dense, low-dimensional representations.
These so called \emph{embeddings} are later used in downstream prediction tasks. 
Given a vector $\Vec{p}\in \mathbb{R}^N$, a decoder function $h$ will try to reconstruct the original vector from the embedding $\Vec{e}\in\mathbb{R}^M$ calculated by the encoder function $g$, such that $\min\Arrowvert \Vec{p} - h(\Vec{e}) \Arrowvert_2^2$ and $M \ll N$. 

\subsection{Prediction and Propagation}
\label{sec:approach_prediction_propagation}

\begin{figure*}
    \centering
    \includegraphics[width=\textwidth]{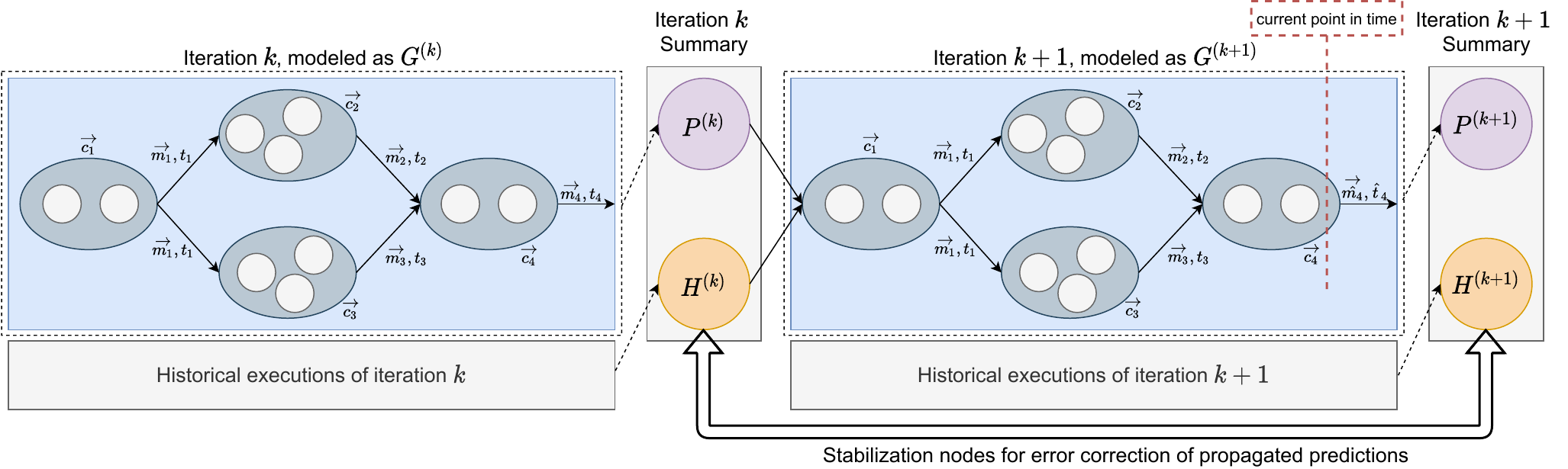}
    \caption{The general idea of Enel. The task sets of each iteration are modeled as graph, and predictions of runtimes and metric vectors are propagated through the graph. We further establish an explicit relationship between multiple iteration graphs by using summary nodes of an iteration, both of the current iteration and its historical executions, as predecessor nodes of the root node(s) of the respective successor iteration graph.}
    \label{fig:approach_enel_approach}
\end{figure*}

For rescaling decisions, we mainly focus on three sources of information that can be gathered for each node $v_i\in V(G^{(k)})$:
\begin{itemize}
    \item \emph{Scale-out}, i.e. the number of workers available at the start and at the end of the respective set of parallel tasks, i.e. $a_i$ and $z_i$. We from here on use the enriched vector representations $\Vec{a}_i, \Vec{z}_i \in \mathbb{R}^3$ altered from the parametric model discussed in~\cite{Venkataraman2016}, with e.g. $\Vec{a}_i$ derivable as $\Vec{a}_i=[1-\frac{1}{a_i}, \log(a_i), a_i)]^\top$.
    \item \emph{Metrics}, which compactly describe the state of the execution environment and the special characteristics of individual sets of parallel tasks, denoted as $\Vec{m}_i\in \mathbb{R}^+$. Examples are CPU or memory utilization.
    \item \emph{Context properties}, which capture the overall context of the job execution and thus allow for differentiating between related contexts. We compute a context vector $\Vec{c}_i\in \mathbb{R}^{3M}$, which is the result of a vector concatenation $\Vec{c}_i = \Vec{u}_i \concat \Vec{v}_i \concat \Vec{w}_i$, with $\Vec{u}_i, \Vec{v}_i, \Vec{w}_i \in \mathbb{R}^M$. The latter denote mean vectors of context embeddings computed in~\autoref{sec:approach_context_encoding}, where $\Vec{u}_i$ covers all always available properties (e.g. job signature), $\Vec{v}_i$ accounts for properties which are not necessarily uniformly recorded (e.g. software versions), and $\Vec{w}_i$ is the mean vector of all properties unique to the respective set of parallel tasks (e.g. number of tasks, attempt ID).
\end{itemize}
These information will be used to compute messages and propagate them through the graph.
We design multiple functions to achieve our goal of graph-based runtime prediction, which we then use to derive reasonable rescaling recommendations. 
Given a node $v_i\in V(G^{(k)})$, we predict its rescaling \emph{overhead} using a function $f_1: \mathbb{R}^+\rightarrow \mathbb{R}$ that takes into account the enclosing context, the metrics recorded for this node, the start and end scale-out, and the fraction of time the node spent in different scale-outs. Formally, we write
\begin{equation}
\hat{o}_i = f_1(\Vec{c}_i, \Vec{m}_i, \Vec{a}_i, \Vec{z}_i, r_i),
\end{equation}
where $\hat{o}_i\in \mathbb{R}$ denotes the predicted overhead, and $r_i$ is the aforementioned fraction of time. Taking the overhead into account, e.g. due to changed data I/O reflected in captured metrics, we subsequently predict the node \emph{runtime} with
\begin{equation}
\hat{t}_i = f_2(\Vec{c}_i, \Vec{m}_i, \Vec{z}_i, \hat{o}_i),
\end{equation}
using a function $f_2: \mathbb{R}^+\rightarrow \mathbb{R}$, with $\hat{t}_i$ being the predicted runtime. Note that $f_2$ only considers the end scale-out, but incorporates the predicted overhead to accurately model the runtime. While $\hat{t}_i$ is the predicted runtime for a particular node, we also compute the total prospective runtime up to a certain node by accumulating the propagated predictions. Thus, for any node $v_i$, the accumulated runtime is established with
\begin{equation}
\hat{tt}_i = \hat{t}_i + \max_{j \in \altmathcal{N}(i)} \hat{tt}_j.
\end{equation}
Consequently, for the last node in a graph, $\hat{tt}_i$ is also the predicted total runtime of the graph.

In order to predict the remaining runtime of a running dataflow job, we need to ensure that the respective function inputs reflect the current job execution as good as possible. As the enclosing execution context is static by definition after a single dataflow job execution, the start and end scale-out for future iterations are ideally equal (i.e. we assume a non-changing scale-out after we have adapted it), an important remaining thing we need to take care of are \emph{metrics}. Thus, we aim at learning the metrics of a node based on its context, scale-out, and the metrics of its predecessor nodes. This would allow us to predict metrics for nodes and propagate these through the graph in order to effectively calculate the desired runtimes. 
In a first step, we aim at learning the relevance of predecessor nodes and their metrics in light of the enclosing context. Inspired by~\cite{Brody2021}, we write $\Vec{x}_i = \Vec{a}_i \concat \Vec{c}_i\concat \Vec{z}_i$ and compute

\begin{equation}
|e_{ij}| = \frac{\text{exp}\left(\Vec{a}^\top\sigma\left(f_3(\Vec{x}_i, \Vec{x}_j)\right)\right)}{
\sum\limits_{k\in \altmathcal{N}(i)}\text{exp}\left(\Vec{a}^\top\sigma\left(f_3(\Vec{x}_i, \Vec{x}_k)\right)\right)},
\end{equation}

i.e. the edge weights, where $f_3: \mathbb{R}^+ \rightarrow \mathbb{R}^+$ is a function that transforms the respective context vectors, $\sigma$ is a non-linear activation, and $\Vec{a}$ is a learnable attention weight vector. 
The scalar result is subsequently normalized using softmax, such that the sum of all edge weights for incoming edges of a node equals one. Eventually, metrics are predicted with
\begin{equation}
\Vec{\hat{m}}_i = \sum_{j\in \altmathcal{N}(i)} |e_{ij}| \cdot f_4(f_3(\Vec{x}_i, \Vec{x}_j), \Vec{m}_j),
\end{equation}
where $f_4: \mathbb{R}^+ \rightarrow \mathbb{R}^+$ is a function that transforms predecessor metrics given the enclosing context. At last, we compute $\Vec{\hat{m}}_i$ as a weighted average of vectors.
We are thus estimating and propagating the metric vectors of successor nodes for which actual metrics are not necessarily available yet. 
The predictions can in turn be used to estimate the corresponding runtimes, which effectively allows us to make predictions for remaining iterations of a respective dataflow job.

In a last step, we explicitly establish a relationship between multiple iteration graphs with the use of summary nodes. 
For a graph $G^{(k)}$, the corresponding summary nodes $P^{(k)}$ and $H^{(k)}$ are installed as predecessor nodes to the respective root node(s) of the next graph $G^{(k+1)}$. 
The summary node $P^{(k)}$ belongs to the component $k$ of the current dataflow job execution, and carries information about the start and end scale-out of the component, and mean vectors of the respective context and metric vectors of graph nodes. The summary node $H^{(k)}$ acts as a historical reference point, where the previously mentioned node information are each averaged across the $\beta$ most similar historical summary nodes of the target component. These $\beta$ nodes are selected based on scale-out proximity. 
The purpose of $H^{(k)}$ is to stabilize the inference process during the propagation of predictions. The summary nodes are only considered for the metric vector prediction.

\section{Architecture}
\label{sec:architecture}

Enel is devised as a blackbox approach, i.e. it is designed as an external component that does not make assumptions about the availability of specific statistics or monitoring data. 
A user can decide which information to use for representing the job execution context and runtime statistics. 
Our approach only requires information about the scale-out at any particular point in time and the fraction of time spent in either start scale-out or end scale-out (e.g. for a set of tasks), which can be fairly assumed to be present in each dataflow system. 

\subsection{Interaction and Usage}
Given a distributed dataflow job, the idea follows that a respective dataflow system can consult our external service and receive a scale-out recommendation for a user-specified runtime target.
Upon each request, Enel tunes a pre-trained model for the targeted job with the most recent runtime information.
It then constructs the remaining component graphs using the static component characteristics gathered from historical job executions, and eventually uses the tuned model for inference on the connected component graphs. 
In this process, predictions are made and pushed through the graphs, until the final component is reached. 
These operations are repeated for any valid scale-out over a defined range and the accumulated runtime predictions are then used to select the configurations which best comply with the runtime targets.
If used for finding a good initial resource allocation, Enel adheres to this process with slight modifications. As the first component of a graph has no predecessor nodes which could be exploited, Enel uses a simple runtime model, namely Bell~\cite{Thamsen2016}, on the historical data of the first graph component, and adds the predicted values to its own given the remaining graph components.

\subsection{Prototypical Implementation}

For our experiments, we developed our prototype with a lightweight web service implemented with Python.
We choose Apache Spark as the distributed dataflow system and schedule Spark applications using the Kubernetes resource manager. 
In this scenario, we implement a listener within Spark that gathers runtime statistics and facilitates communication with our service. 
Regarding metrics used in our optimization process, five were selected directly from within the Spark listener, namely CPU utilization, Shuffle R/W, Data I/O, the fraction of time spent in garbage collection, and the ratio of memory spilled to disk to peak execution memory. 
Enel uses these metrics together with context properties to predict the remaining runtime and, if required, recommends an optimized scale-out.
This scale-out recommendation is then used within our Spark listener to renegotiate resources with Kubernetes.

\subsection{Prediction Model}
For modeling we employ soft computing, i.e. we realize each of our differentiable functions $f_1, f_2, f_3$, and $f_4$ as a two-layer feed-forward neural network. 
The latter are good function estimators with two layers being sufficient to distinguish data that is not linearly separable. 
Operating on graphs and propagating messages eventually makes our model a spatial graph neural network, as we are defining our calculations in the vertex domain by leveraging the graph structure and aggregating node information.
With a total of 5155 learnable parameters, our model allows for training even using a CPU.

\section{Evaluation}
\label{sec:evaluation}
This section presents our infrastructure setup and experiment design, followed by a discussion of the obtained results. 
Further model implementation details and technical information are provided in our repository\footnote{https://github.com/dos-group/enel-experiments}.

\subsection{Infrastructure Setup}
\label{sec:evaluation_infrastructure_setup}
All experiments were done on a multi-tenant Kubernetes cluster spanned over 50 machines, which makes each of our scheduled workloads compete with other workloads in the cluster about resources. Of the machines used for deploying the cluster, 42 were also used to roll out a Hadoop Distributed File System (HDFS), where the datasets and spark jobs were located. Each Spark application scheduled within the Kubernetes cluster used a scale-out range from 4 to 36 Spark executors. Further details are reported in~\autoref{tbl:cluster_specs}.

\begin{table}
\centering
\caption{Hardware \& Software Specifications}
    \begin{tabular}[t]{lr}
        \toprule
        \multicolumn{2}{c}{\textbf{\underline{Hardware}}}\\
        CPU, vCores & Intel(R) Xeon(R) CPU @ 3.30 GHz, 8\\
        Memory & 16 GB RAM\\
        Network & connected by a single 1 GB switch\\
        \midrule
        \multicolumn{2}{c}{\textbf{\underline{Software}}}\\
        Linux & Kernel 4.15.0\\
        Others & Python 3.8.0, Kubernetes 1.18.10\\
        & Hadoop 2.8.3, Scala 2.12.11\\ &
        Spark-Operator\footnote{https://github.com/GoogleCloudPlatform/spark-on-k8s-operator, accessed 2021-06-08} 1.1.3, Spark 3.1.1\\
        & PyTorch 1.8.0, PyTorch Geometric 1.7.2\\
        \midrule
        \multicolumn{2}{c}{\textbf{\underline{Spark}}}\\
        Driver CPU & 6 cores\\
        Driver Memory & 10240 MB (+ 2048 MB)\\
        Executor CPU & 6 cores\\
        Executor Memory & 10240 MB (+ 2048 MB)\\
        \bottomrule
    \end{tabular}
\label{tbl:cluster_specs}
\end{table}

\begin{table}
\centering
\caption{Overview of Benchmark Jobs}
    \begin{tabular}[t]{c|c|c|l}
        \textbf{Job}&\textbf{Dataset}&\textbf{Input Size}&\multicolumn{1}{c}{\textbf{Parameters}}\\
        \toprule
        LR & Multiclass & 27 GB & 20 iterations\\ 
        \midrule
        MPC & Multiclass & 27 GB & \begin{tabular}[c]{@{}l@{}}20 iterations, 4 layers with\\200-100-50-3 perceptrons\end{tabular}\\
        \midrule
        K-Means & Points & 48 GB & 10 iterations, 8 clusters\\
        \midrule
        GBT & Vandermonde & 35 GB & \begin{tabular}[c]{@{}l@{}}10 iterations,\\“Regression” configuration\end{tabular}\\
        \midrule
    \end{tabular}
\label{tbl:benchmark_jobs}
\end{table}

\subsection{Experiment Design}
\label{sec:evaluation_experiment_design}
In order to assess the applicability of our approach, we conduct a series of experiments, in which we compare ourselves to related works and simulate the prediction robustness by injecting synthetic failures during spark application executions. 
We are primarily interested in investigating if and how good our single model, which is reusable due to its context-awareness, performs in comparison to related work that utilizes an ensemble of specialized models.

\subsubsection{Jobs}
We used four Spark jobs as benchmarks, namely Logistic Regression (LR), Multilayer Perceptron Classifier (MPC), K-Means, and Gradient Boosted Trees (GBT). \autoref{tbl:benchmark_jobs} shows the jobs and the respective input parameters. The implementation are based on Spark MLLib\footnote{http://spark.apache.org/mllib/, accessed 2021-06-08}, a library for implementing distributed machine learning algorithms. 

\subsubsection{Datasets}
We used three different datasets for the benchmark jobs. All datasets were generated synthetically.
\begin{itemize}
    \item Multiclass: A classification dataset with 3 classes and 200 features. The dataset was generated using the classification generator from scikit-learn\footnote{https://scikit-learn.org/, accessed 2021-06-08}.
    \item Vandermonde:  A regression dataset with 19 features. The dataset was generated using our own generator by explicitly computing the Vandermonde matrix. For this, data points were randomly generated following a polynomial of degree 18 with added Gaussian noise.
    \item Points: A two-dimensional dataset with points sampled from a Gaussian Mixture Model of eight normal distributions with random cluster centers and equal variances.
\end{itemize}

\subsubsection{Objective: Prediction Performance}
For each dataflow job, we first perform an initial ten runs without dynamic scaling. 
The execution data of these ten runs are gathered and later used to train models for initial resource allocation and dynamic scaling. 
For dynamic scaling, we compare our approach to our own previous work \emph{Ellis}~\cite{Thamsen2017}, as it is most related.
For initial resource allocation, we rely on the method presented in Ellis to guarantee an equal and fair starting point for both dynamic scaling approaches.
After initial model fitting, we conduct various adaptive runs and hereby investigate the performance of our approach. 
While Ellis is by design fitting and using a new set of models for every new run, we train with Enel a new model from scratch after every fifth run and fine-tune this model then on each of the subsequent five runs. 

\subsubsection{Objective: Prediction Robustness}
In order to test the reliability of our method in the presence of partial failures, i.e. the loss of executors while a job is executing, we created a failure injector to inject failures into the running system. Our failure experiments were structured in such a way that they executed across the full time duration of each job execution, where a single failure was injected at any random second within a 90 second interval as long as at least 4 Spark executors were available. Injections resulted in the immediate termination of Spark executor pods, which in turn required Spark, if configured accordingly, to start up new or additional executors to cover for that. 
We conduct multiple runs which are subject to failure injections, distributed across two phases and interrupted by a handful of normal runs.

\subsubsection{Evaluation Metrics}
We assess the performance of our approach using multiple metrics. First, we utilize metrics introduced in~\cite{Thamsen2017}, namely CVC (runtime constraint violation count) and CVS (runtime constraint violation sum). Moreover, we investigate the time to train required for our approach.

\begin{table*}
\centering
\caption{Evolution of prediction performance over runs}
\begin{threeparttable}
\begin{tabular}{@{}c|c||cc||cc||cc||cc||cc@{}}
\multicolumn{2}{c}{} & \multicolumn{2}{c||}{\textbf{Runs 11-22}} & \multicolumn{2}{c||}{\textbf{Runs 22-33}} & \multicolumn{2}{c||}{\textbf{Runs 33-44}} & \multicolumn{2}{c||}{\textbf{Runs 44-55}} & \multicolumn{2}{c}{\textbf{Runs 55-65}} \\ \cmidrule{3-12}
\multicolumn{2}{c}{} & $\bar{x}$\ \tnote{1}       & $\tilde{x}$\ \tnote{2}         &           $\bar{x}$       & $\tilde{x}$         &           $\bar{x}$       & $\tilde{x}$  & $\bar{x}$       & $\tilde{x}$ & $\bar{x}$       & $\tilde{x}$ \\ \midrule
\multirow{2}{*}{\textbf{LR}} & CVC & $0.55$ & $1.00$ & $0.36$ & $0.00$ & $0.73$ & $1.00$ & $1.00$ & $1.00$ & $0.91$ & $\mathbf{1.00}$  \\
                  & CVS & $0.33$m & $0.11$m& $0.12$m & $0.00$m& $1.40$m & $0.31$m& $1.44$m & $1.35$m& $1.58$m & $\mathbf{0.92}$m   \\ \midrule
\multirow{2}{*}{\textbf{MPC}}  & CVC & $0.73$ & $1.00$ & $0.27$ & $0.00$ & $0.45$ & $0.00$ & $0.82$ & $1.00$ & $0.27$ & $\mathbf{0.00}$  \\
                  &  CVS & $2.41$m & $1.70$m& $1.12$m & $0.00$m& $1.95$m & $0.00$m& $2.46$m & $1.22$m& $1.23$m & $\mathbf{0.00}$m \\ \midrule
\multirow{2}{*}{\textbf{K-Means}}  & CVC & $0.55$ & $1.00$ & $0.27$ & $0.00$ & $0.00$ & $0.00$ & $0.00$ & $0.00$ & $0.00$ & $\mathbf{0.00}$    \\
                  & CVS & $5.89$m & $0.60$m& $4.03$m & $0.00$m& $0.00$m & $0.00$m& $0.00$m & $0.00$m& $0.00$m & $\mathbf{0.00}$m   \\ \midrule
\multirow{2}{*}{\textbf{GBT}}  &  CVC & $0.36$ & $0.00$ & $0.36$ & $0.00$ & $0.82$ & $1.00$ & $0.73$ & $1.00$ & $0.27$ & $\mathbf{0.00}$  \\
                  & CVS & $0.55$m & $0.00$m& $0.85$m & $0.00$m& $5.21$m & $2.05$m& $2.85$m & $1.04$m& $0.27$m & $\mathbf{0.00}$m
\end{tabular}
\begin{tablenotes}
        \item[1] The \emph{mean} value of the metric over the respective runs.
        
        \item[2] The \emph{median} value of the metric over the respective runs.
    \end{tablenotes}
\end{threeparttable}
\label{tbl:evaluation_complete_experiment_evolution}
\end{table*}

\begin{figure}
    \centering
    \includegraphics[width=\columnwidth]{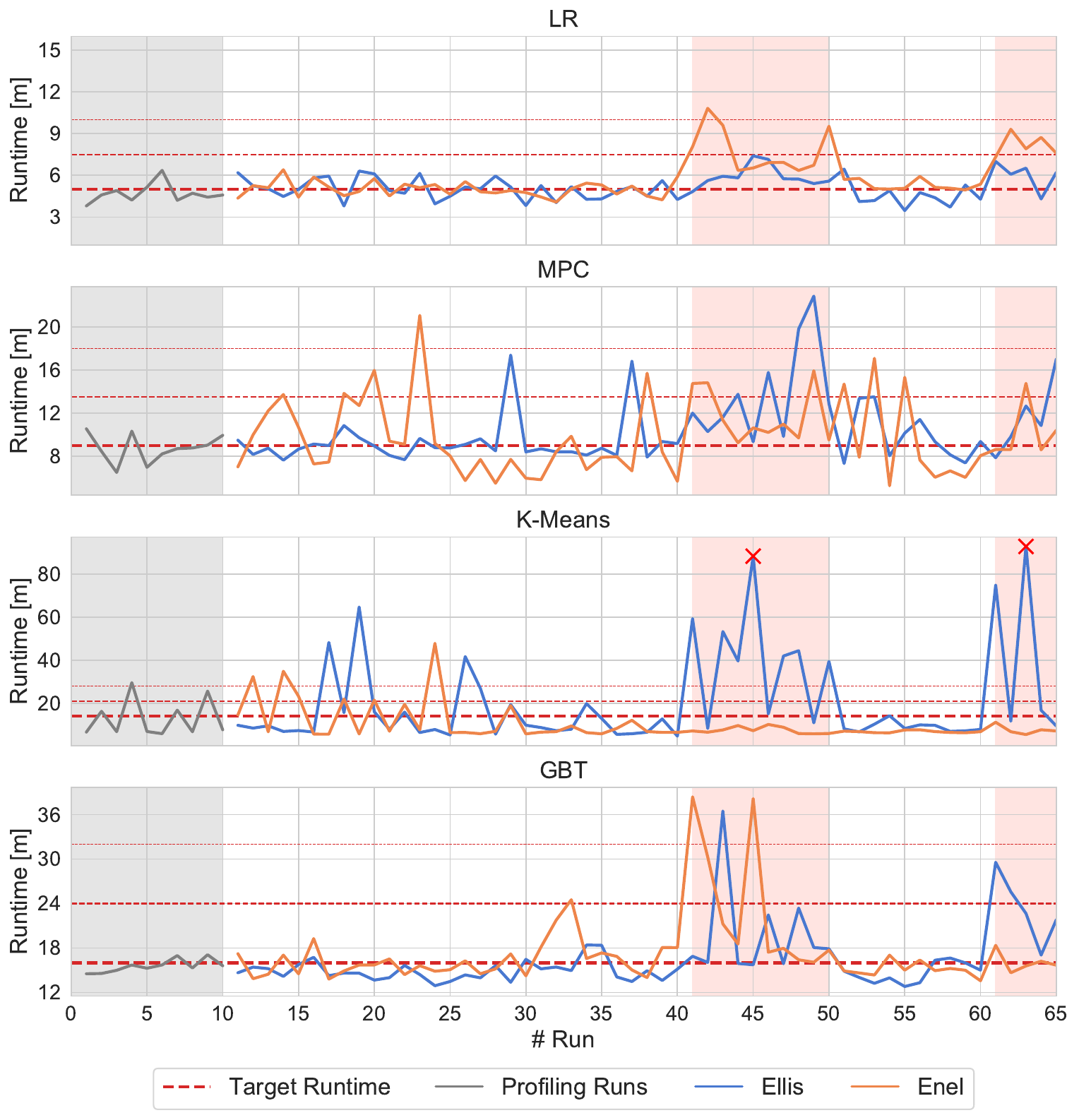}
    \caption{Initial profiling (highlighted in grey), followed by alternating phases of normal runs and anomalous runs (highlighted in red). Our graph model Enel is improving over time and shows a certain robustness against failures.}
    \label{fig:evaluation_complete_experiment_runtime}
\end{figure}

\begin{figure}
    \centering
    \includegraphics[width=\columnwidth]{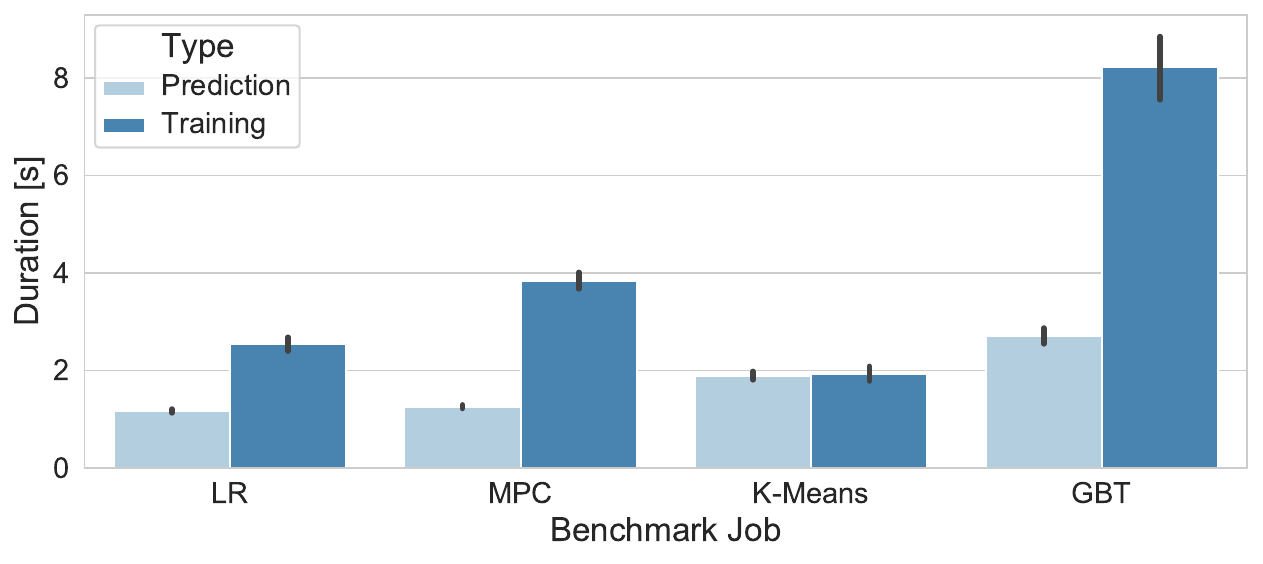}
    \caption{Time in seconds required for fine-tuning an Enel model and eventually using it for predictions. The variance for some jobs stems from the fact that training takes differently long dependent on the number of graphs to process.}
    \label{fig:evaluation_duration_fit_predict}
\end{figure}

\subsection{Results}
\label{sec:evaluation_results}

The general prediction performance over all runs is illustrated in~\autoref{fig:evaluation_complete_experiment_runtime}. 
It is observable that Enel, due to its complexity and the limited training data, is initially often volatile in its prediction performance. 
This is especially evident for runs with MPC or K-Means.
Over time though, the predictions tend to stabilize and converge, with the exception of single outliers. 
In~\autoref{tbl:evaluation_complete_experiment_evolution}, we report mean and median values of CVC and CVS for each algorithm over multiple ranges. 
The results support our previous finding of improved prediction performance over time.
In general, the predicted runtime is many times only slightly off and regularly below the respective formulated runtime target. 
From this we conclude that achieving results comparable to state-of-the-art solutions is possible given a certain amount of training data to work with.
This is interesting, as we only utilize a single model operating on individual graphs and perform training only occasionally, whereas Ellis trains a new set of models after every run, with each model being specialized for a concrete job component. 
From this we can deduce that the enclosing execution contexts were sufficiently represented using descriptive properties, and that our model was able to make context-aware predictions with the computed embeddings.
Moreover, learning the relationships between graphs appears to be beneficial. 

In terms of prediction robustness, e.g. in the presence of node failures, Enel reveals a certain stability, where especially the results on K-Means are worth mentioning. 
In case of initial problems during anomalous runs, the prediction error is significantly corrected with the next scheduled retraining, as can be seen for GBT after iteration 45. For LR, the decisions of Enel tend to lead to a slightly longer runtime, however, as the runtime target is in general quite low, the magnitude of violation is tolerable.
Meanwhile, the comparative method Ellis faces difficulties reacting to certain failures, which maximizes for MPC and K-Means. The problems persist also for the second phase of anomalous runs. As indicated by the red crosses, certain runs even had to be terminated manually because a timely application end was not realistic.
We thus conclude that Enel is able to react more appropriately to failures due to its consideration of runtime statistics. 

Lastly, we discuss the required time for training and prediction. 
\autoref{fig:evaluation_duration_fit_predict} summarizes the measured times for each job class. It can be seen that inference is a fast and robust process, whereas the fine-tuning of a model depends on the number of provided graphs. 
As a consequence, GBT requires more time as the job is internally decomposed into many components. 
Either way, we argue that the cost of a few seconds is acceptable given that dataflow jobs usually run for longer periods of times. 
Note that we did not list the time required for the initial training of the models, as this is a task that can be scheduled for execution periodically.

\section{Conclusion}
\label{sec:conclusion}
This paper presented \emph{Enel}, a novel dynamic scaling approach for distributed dataflows. 
It explicitly incorporates the graph structure of distributed dataflow jobs as well as descriptive properties of job execution contexts to develop a better understanding of how scale-out behaviors are impacted during execution. 
At any point in time, Enel is able to estimate the remaining runtime of a targeted job and then compares it to projected runtime targets. 
The resource allocation is then accordingly adapted to at best comply with existing constraints. 
With the consideration of descriptive context properties, Enel can be reused across multiple iterations of an iterative dataflow job instead of training individual models. Moreover, it is a black-box approach and thus can be used with different resource managers and for different dataflow systems.

We implemented Enel as a graph neural network with multiple task-specific components, which for instance realized the prediction of runtimes and runtime statistics.
Our experiments showed that as more data becomes available, Enel is able to improve its predictions and thus achieves comparable results to related works and additionally is able to handle failures in a more robust manner.
The advantage of our approach is especially significant for iterative dataflow jobs which incorporate small changes in terms of configurations or inputs.
This is as a result of the reusability of our model and thus knowledge from previous related executions is recycled which saves on resource and energy usage. 

In the future, we want to further investigate possibilities of exploiting the dataflow graph information for optimized resource allocation, dynamic scaling, and scheduling of multiple dataflow jobs. 
As this is a common scenario in shared clusters, sophisticated solutions are required and desired.

\section*{Acknowledgments}
This work has been supported through grants by the German Federal Ministry of Education and Research (BMBF) as BIFOLD (funding mark 01IS18025A) and WaterGridSense 4.0 (funding mark 02WIK1475D).

\bibliographystyle{IEEEtran}
\bibliography{bib}

\end{document}